\documentstyle[preprint,aps]{revtex}
\begin{document}

\preprint{BGU PH-96/04}
\draft

\title{Extensible Black Hole Embeddings\\
for Apparently Forbidden Periodicities}

\author{Aharon Davidson\thanks{davidson@bgumail.bgu.ac.il} and 
\addtocounter{footnote}{3}
Uzi Paz\thanks{uzipaz@bgumail.bgu.ac.il}}
\address{ Physics Department,
Ben-Gurion University of the Negev,
Beer-Sheva 84105, Israel}

\maketitle

\begin{abstract}

Imposing extendibility on Kasner-Fronsdal black hole local
isometric embedding is equivalent to removing conic singularities
in Kruskal representation.
Allowing for globally non-trivial (living in $M_{5}\times S_{1}$)
embeddings, parameterized by $k$, extendibility can be achieved for
apparently forbidden frequencies $\omega_{1}(k)\le\omega (k)\le
\omega_{2}(k)$.
The Hawking-Gibbons limit, $\displaystyle{\omega_{1,2}(0)=
{1\over{4M}}}$ for Schwarzschild geometry, is respected.
The corresponding Kruskal sheets are viewed as slices in some
Kaluza-Klein background. 
Euclidean $k$ discreteness, dictated by imaginary time periodicity,
is correlated with twistor flux quantization.

\end{abstract}

\pacs{PACS numbers: 02.40.-k, 04.70.-s, 04.70.Dy, 11.10.Kk}

\pagebreak

Local isometric embedding \onlinecite{exact} of a curved
$d$-dimensional manifold, within some parent $D$--dimensional
flat space-time, has been traditionally invoked to classify
\onlinecite{NR} the variety of general relativity solutions.
The embedding is fully characterized by its induced metric,
twistor Yang-Mills gauge field and the extrinsic curvature
which, on consistency grounds, are subject to Gauss, Codazzi,
and Ricci equations.
Interesting attempts $(i)$ to interpret the embedding
functions as alternative canonical variables for gravity
\onlinecite{RT}, $(ii)$ to relate the normal space symmetries
with the electro/nuclear interactions \onlinecite{RMP}, and
$(iii)$ to view our world as confined to a ($3+1$)--dimensional 
membrane by some potential well \onlinecite {RS}, are worth
recalling.
Depending on the differentiable nature of the embedding
functions, $D \le \frac{1}{2} d(d+1)$ for analytic embeddings
\onlinecite{JC}, whereas $D\le\frac{1}{2}d(d+3)$ if only
integrability is required \onlinecite{G}.
Of particular interest for us, however, are the ($3+1$)--dimensional
radially symmetric solutions; they fall categorically into the
$D=6$ embedding class.
The latter fact was already known to Kasner \onlinecite{K} who 
presented a $(4+2)$--dimensional embedding, non-causal and
non-extendible though, of the exterior Schwarzschild geometry.
However, only the Fronsdal $(5+1)$--dimensional embedding
\onlinecite{F} has the advantage of being one-to-one correlated
with the Kruskal-Szekeres analysis \onlinecite{KS}.
As far as black holes geometries are concerned, the removal of
apparent horizon singularities in the Lorentzian Kruskal
representation, which is fully equivalent to discharging conic
singularities in the Euclidean world by means of Hawking-Gibbons 
\onlinecite{HG} periodicity, can be translated into imposing
extendibility in the Kasner-Fronsdal approach.
In this paper, following a pedagogical introduction, we present 
extensible Schwarzschild embeddings for apparently forbidden
periodicities.

\bigskip
To be more specific, but keeping a certain amount of generality,
consider the radially symmetric $4$--metric
\begin{equation}
	ds^2 = -A(r)dt^2 + {1 \over {B(r)}}dr^2 + r^2d\Omega,
	\label{metric}
\end{equation}
where $d\Omega\equiv d\vartheta^2+\sin^2\vartheta d\varphi^2$.
For this metric to describe a black hole, $\sqrt{AB}$ must well
behave near the critical radius where $A(r_{h}) = 0$.
Indeed, invoking a set of Kruskal coordinates, namely 
$v = C(r)\sinh \omega t$ and
$u = C(r)\cosh \omega t$,
one is led to the compelling requirement that
${\displaystyle A\exp(-2\omega\int{{{dr}\over{\sqrt{AB}}}})}$
must reach a non-zero finite value at the horizon.
This is to assure a singularity-free scale for the light-cone
combination $(-dv^{2}+du^{2})$.
In turn, the parameter $\omega$ gets fixed
\begin{equation}
	\omega =\left.{{1\over2}\sqrt{{B\over A}}{{dA}\over{dr}}}
	\right|_{r=r_h}.
	\label{omega}
\end{equation}
The would be imaginary time periodicity ${\displaystyle
\frac{2\pi}{\omega}}$ is the Hawking-Gibbons \onlinecite{HG} key to
Bekenstein-Hawking \onlinecite{thermo} black hole thermodynamics.

\bigskip
Alternatively, one may consider the embedding of the above
$4$--metric in $M_{6}$ with flat metric
\begin{equation}
	ds^2 = -dy_0^2+\sum\limits_{n=1}^5 {dy_n^2}.
\end{equation}
Apart from the usual assignments $y_1=r\cos \vartheta$,
$y_2=r\sin \vartheta \cos \varphi$, and $y_3=r\sin \vartheta
\sin \varphi$, which define the radial marker, one further
introduces
\begin{eqnarray}
	y_0 &=& f(r)\sinh \omega t,\nonumber \\
	y_4 &=& f(r)\cosh \omega t, \\
	\label{ansatz}
	y_5 &=& g(r).\nonumber	
\end{eqnarray}
These are supposed to cover the (say) $\displaystyle{\left|{{{y_0}
\over{y_4}}}\right|\le{1}}$ section of the $4$-manifold
characterized by $\displaystyle{y_4^2-y_0^2={A\over{\omega^2}}}$.
After some algebra we arrive at
\begin{equation}
	\left({{{dg}\over{dr}}}\right)^2 =-1+
	{1\over B}-{1\over{4\omega^2A}}
	\left({{{dA}\over{dr}}}\right)^2,	
\end{equation}
noticing that it is precisely that $\omega$ given by
eq.(\ref{omega}) for which $\displaystyle{{dg}\over{dr}}$ remains
finite at the horizon.
This establishes the correspondence between the Kruskal
and the Fronsdal schemes.

\bigskip
For pedagogical reasons, let us focus attention on the
Schwarzschild geometry, specified by $\displaystyle{A(r)=B(r)
=1-{{2m}\over{r}}}$.
The crucial point then is the interplay of the two zeroes
of $\displaystyle{\sqrt{AB}{{dg}\over{dr}}}$.
One infers that,

\noindent 
($i$) For $\displaystyle{\omega \ge {1\over{4m}}}$, the
embedding does not cover the interior strip $\displaystyle
{\left({{m\over{2\omega^2}}}\right)^{{1\over 3}} < r < 2m}$,
whereas

\noindent
($ii$) For $\displaystyle{\omega \le {1\over{4m}}}$, the
embedding does not cover the exterior strip $\displaystyle
{2m < r < \left({{m\over{2\omega^2}}}\right)^{{1\over 3}}}$.

\noindent
The conclusion being that Fronsdal extendibility requires
\begin{equation}
	\left({1-{{2m}\over{r}}}\right)\left({1-{m\over{2
	\omega^2r^3}}}\right)\ge{0}
	\label{extend}
\end{equation}
for all $0<r<\infty$, and thus can be achieved only provided
$\displaystyle{\omega ={1\over{4m}}}$, recognized as the
Kruskal value.
Appreciating the role played by inequality (\ref{extend}),
we note that its direction would have been harmfully reversed
had we considered a time-like $y_{5}$.

\bigskip
We now claim that eq.(\ref{ansatz}) is in fact \underline{not}
the most general ansatz capable of giving birth to the radially
symmetric (and hence time independent) metric (\ref{metric}).
Introducing a new parameter $k$, one can still have
\begin{eqnarray}
	y_0 &=& f(r)\sinh [\omega{t}+k\psi(r)],\nonumber \\
	y_4 &=& f(r)\cosh [\omega{t}+k\psi(r)],\\
	y_5 &=& kt+g(r),\nonumber
\end{eqnarray}
with $k=0$ serving as the Fronsdal limit.
We now state, skipping the proof due to length limitations
(to be published elsewhere), that the embedding in hand is
accompanied by the $SO(2)$ twistor vector field of the two
normal directions 
\begin{equation}
	A_t(r)=-k\omega\sqrt{{B\over A}}.
	\label{twist}
\end{equation}
The latter vanishes at the Fronsdal limit, and appears as a
harmless pure gauge configuration for a long list of general
relativity solutions (including Schwarzschild and Reissner-Nordstrom).
Nonetheless, for $k\neq 0$, it may leave non-trivial global
fingerprints upon Euclidization, once imaginary time is
identified with a certain period. 

\bigskip
The functions involved in the embedding procedure are:
\begin{eqnarray}
	f^2 &=& \frac{1}{\omega^{2}}\left(1+k^2-{{2m}\over{r}}\right),
	\nonumber \\
	{{dg}\over{dr}} &=& {1\over{\left({1-{{2m}\over{r}}}\right)}}
	\sqrt{\ },\\
	{{d\psi}\over{dr}} &=& {1\over{\omega{f^2}}}{{dg}\over{dr}},
	\nonumber
\end{eqnarray}
where $\displaystyle{\sqrt{\ }\equiv\sqrt{{{2m}\over{r}}\left[
{k^2+(1-{{2m}\over{r}})(1-{m\over{2\omega^2r^3}})}\right]}}$.
The first thing to notice is that, unlike in the Fronsdal
limit, both $g(r)$ and $\psi(r)$ are now singular (logarithmic
singularity) at $r=2m$.
This is not necessarily a problem, however, as one has still
the option of performing a gauge transformation (that is, a
general coordinate transformation), and blame it for the
singularity induced.
Radial symmetry (and hence time independence by virtue of
Birkhoff theorem) still allows for $t\to{t}+\Lambda(r)$.
Such a shift in $t$ is equivalent to a redefinition of $g(r)$
and $\psi(r)$, namely $g\to{g}+k \Lambda$ and $\displaystyle
{\psi\to{\psi}+{\omega\over{k}}\Lambda}$, under which only the
combination $\displaystyle{g-{{k^2}\over\omega}\psi}$ would
not change.
The fact that such a `gauge-invariant' combination exhibits
no singularity at $r=2m$, as can be seen from $\displaystyle{
\frac{d}{dr}(\psi-\frac{\omega}{k^{2}}g)=-\frac{\sqrt{\ }}{
\omega k^{2}f^{2}}}$, comes with no surprise.

\bigskip
Two tenable gauges offer their services:

\noindent $(i)$ Using $\displaystyle{{{d\Lambda_1}\over{dr}}=
-{{\sqrt{\ }}\over{k\left({1-{{2m}\over{r}}}\right)}}}$, one is
led to the convenient choice $g_{1}=0$ accompanied by $\displaystyle
{\psi_{1}=\psi-\frac{\omega}{k^{2}}g}$, and

\noindent $(ii)$ Using $\displaystyle{{{d\Lambda_2}\over{dr}}=
-{{k\sqrt{\ }}\over{\omega^2f^2\left({1-{{2m}\over{r}}}\right)}}}$,
on the other hand, paying attention to the extra $f^{2}$ in the 
denominator, one obtains $\displaystyle{g_{2}=g-\frac{k^{2}}
{\omega}\psi}$ on the expense of $\psi_{2}=0$.

\noindent Both gauges give rise to non-diagonal Schwarzschild
variants of the Eddington-Finkelstein type.

\bigskip
Witnessing the smooth behavior of the embedding functions
near the horizon $r_{h}=2m$, we turn attention now to the
apparently critical radius $\displaystyle{r_{c}=\frac{2m}
{1+k^{2}}}$. This is where the matching of the two sections
$\displaystyle{\left|{{{y_0}\over{y_4}}}\right|\le{1}}$ and
$\displaystyle{\left|{{{y_0}\over{y_4}}}\right|\ge{1}}$ of the
manifold characterized by $\displaystyle{y_4^2-y_0^2=\frac{1}{
\omega^{2}}\left(1+k^2-{2m\over{r}}\right)}$ is supposed to
take place.
A closer inspection of the two gauges a priori permissible
is thus in order.
The first gauge offers us the advantage that $f\cosh{k\psi_{1}}$
and $f\sinh{k\psi_{1}}$, the amplitudes of $\sinh{\omega t}$
and $\cosh{\omega t}$, are perfectly regular at $r=r_{c}$.
To be more specific, $f\sim\sqrt{\epsilon}$ whereas $\displaystyle
{k\psi_{1}\sim{-\frac{1}{2}\ln{\epsilon}}}$ as $\epsilon\equiv r-r_{c}
\to{0}$. 
The second gauge offers us nothing but a major drawback: namely,
$g_{2}$ gets (logarithmically) singular at $r=r_{c}$.
The gauge choice is thus clear.

\bigskip
For the embedding to cover any given region of the $4$--dim
manifold, the $\displaystyle{\sqrt{\ }}$ function involved must
stay real in that region.
Consequently, generalizing eq.(\ref{extend}) in a very simple
manner, $k$-extendibility requires
\begin{equation}
	\left({1-{{2m}\over{r}}}\right)\left({1-{m\over{2\omega^2r^3}}}
	\right)+k^2\ge{0}
	\label{kextend}
\end{equation}
to hold for all $0<r<\infty$.
We now argue that if $\omega \le \omega_{1}(k)$, the embedding does
not cover some (exterior) strip, whereas $\omega \ge \omega_{2}(k)$
leaves another (interior) strip without coverage.
While fully respecting the $k=0$ Hawking-Gibbons limit, the door
is widely open now for apparently forbidden black hole frequencies
in the range $\omega_{1}(k)\le\omega\le\omega_{2}(k)$.
The allowed region in the $(\omega ,k)$--plane is depicted
in the enclosed Figure.
$\omega_{1,2}(k)$ are the two roots of
\begin{equation}
	\left({1-{{2m}\over{a}}}\right)\left({1-{m\over{2\omega^2a^3}}}
	\right)+k^2=0,
\end{equation}
where the radius $a(m,\omega)$, for which $\displaystyle{\left(
{1-{{2m}\over{r}}}\right)\left({1-{m\over{2\omega^2r^3}}}\right)}$
is minimal, is given by
\begin{equation}
	a(m,\omega)={{m^{{1\over{3}}}}\over{\omega^{{2\over{3}}}}}
	\left[{\left({\sqrt{1+{1\over{64m^2\omega^2}}}+1}\right)^{{1
	\over{3}}}-\left({\sqrt{1+{1\over{64m^2\omega^2}}}-1}
	\right)^{{1\over{3}}}}\right].	
\end{equation}
In particular, for large $k$, the extremal $\omega_{1,2}(k)$ behave like
\begin{eqnarray}
	\omega_{1}(k) &\simeq& {{3\sqrt 3} \over {64mk}},\\
	\omega_{2}(k) &\simeq& {{4k^3} \over {3\sqrt 3m}}.
\end{eqnarray}

\bigskip
In search of proper `dispersion relations', the naive candidate
family is of course $\omega(\xi k)$,
conveniently parameterized by the continuous parameter
$-1\le\xi\le{1}$, and given as the solution of
\begin{equation}
	\left({1-{{2m}\over{a}}}\right)\left({1-{m\over{2\omega^2a^3}}}
	\right)+\xi^{2}k^{2}=0.
\end{equation}
Near the Hawking-Gibbons limit, that is for small enough $k$, one 
derives
\begin{equation}
	\omega (\xi k)\simeq{1\over{4m}}(1+\sqrt{3}\xi k).
\end{equation}

\bigskip
The global aspects of the $k$--embedding are next.
The fundamental role played by the hyperbolic functions of
$\omega t$ in the Kruskal and the Fronsdal schemes is very much
established by now.
But here, the situation appears to be a bit more complicated,
due to the fact that a linear function of $t$, namely $y_{5}=kt$
(using the preferred $\Lambda_{1}$--gauge), is floating around
as well.
We first infer, recalling that the argument of the hyperbolic
functions is $\omega{t}+k\psi_{1}(r)$, that when going Euclidean,
$t\to{i\tau}$ must be accompanied by $k\to{-i\kappa}$ (and also
by $\xi\to{i\zeta}$).
In turn, and this is a central point, $y_{5}\to{y_{5}}$ without
a change in signature.
A potential problem then arises: Imaginary time periodicity is
violated in principle, unless of course the fifth dimension acts
cooperatively.
In other words, if $\tau$-periodicity is important to us (and we 
believe quantum mechanics is rather important), $y_{5}$ better be
a closed coordinate, and this must be the case at the
Lorentzian level as well.
This is why the embedding space-time must have the topology of
$M_{5}\times S_{1}$ (to be contrasted with Fronsdal's $M_{6}$),
thereby establishing the arena for the linear function of $\tau$
to play its non-trivial global role.
We remark in passing that the Schwarzschild metric is not
a plane-wave metric, and hence its global $(1+N)$-embedding
in not Penrose restricted $\onlinecite{RMP}$.

\bigskip
Now, the $\tau$-periodicity of $\displaystyle{\frac{2\pi}{
\omega}}$ must be in accord with the underlying topology, but
this can only be discretely satisfied, leading to
\begin{equation}
	\kappa_{n}=n\omega{R},
	\label{kappa}
\end{equation}
with $R$ denoting the radius of the fifth dimension.
The latter quantization condition has a rather interesting
6--dimensional interpretation.
Recalling the attached twistor vector potential ($\ref{twist}$),
and noticing that
\begin{equation}
	\int{A_\mu}dx^\mu\to\kappa\omega\oint{d\tau}=2\pi\kappa, 
\end{equation}
one realizes that (\ref{kappa}) is nothing but magnetic flux
quantization in disguise (in Kaluza-Klein-like units, with
$\displaystyle{\frac{1}{2\pi\omega R}}$ serving as the twist
electric charge).

\bigskip
To complete the correspondence between the Kruskal removal of
conic singularities and the Fronsdal extendibility, we are after
the so-called $k$--generalization of the original Kruskal scheme.
Let our starting point be the $5$-geometry
\begin{equation}
	ds_{5}^{2}=dx_{5}^{2}+ds_{4}^{2},
\end{equation}
where $x_{5}$ is a compactified (a la Kaluza-Klein) fifth
dimension, and the $4$-metric takes the form
\begin{equation}
	ds_4^2 \equiv -\left({1+k^2-{{2m}\over r}}\right)dt^{2}+
	\left[{1+{{m^2}\over{\left({1+k^2-{{2m}
	\over r}}\right)\omega^2 r^4}}}\right]dr^2+r^2 d\Omega.
	\label{Kruskal'}
\end{equation}
The above carefully designed $4$--metric exhibits a major feature.
Namely, as can be verified by means of eq.($\ref{omega}$), this
metric is Kruskalizable for any arbitrary prescribed $\omega$.
There is no mystery about this; $ds_{4}^2$ has the familiar Fronsdal
embedding in $M_{5}$, that is $ds_{4}^2=-dy_0^2+\sum\limits_{n=1}^4
{dy_n^2}$ (with $n=5$ notably excluded).
Using momentarily Euclidean language, where $t\to{i\tau}$ (and $k
\to{-i\kappa}$), we deal with a torus specified by its periodicities
$\Delta x_{5}=2\pi R$ and $\displaystyle{\Delta\tau=\frac{2\pi}
{\omega}}$.

\bigskip
Consider now a class of $4$--dimensional manifolds which reside
within the given $5$--dimensional space-time, and proceed in steps:

\noindent $(i)$ By arbitrarily assigning $x_{5}(x^{\mu})$, one
induces the $4$--metric $\displaystyle{ds_{4}^{2}+{\left({\frac{
\partial x_{5}}{\partial x^{\mu}}dx_{\mu}}\right)}^{2}}$.
This will generically kill all parent periodicities.

\noindent $(ii)$ Symmetry-wise, one can do better by choosing
$x_{5}(t,r)=at+b(r)$. 
This way, and here we switch again to the Euclidean framework (with 
$a\to -i\alpha$), one may still recover closed lines on the torus.
In particular, if $\alpha_{n}=n\omega R$, the $\displaystyle{\Delta
\tau=\frac{2\pi}{\omega}}$ periodicity stays alive.

\noindent $(iii)$ If it so happens that the induced metric is
locally Schwarzschild, we are done.

\bigskip
The prescription is then clear. 
Given the seed metric (\ref{Kruskal'}), first shift
\begin{equation}
	t\to{\tilde{t}+\frac{k}{\omega}\psi (r)},
\end{equation} 
and then cut out the simple Kaluza-Klein slices
\begin{equation}
 	x_{5}(\tilde{t},r)=k\tilde{t}+ g(r).
\end{equation} 
To import no apparent singularities this way, however, it is
advisable to make contact with our preferred $\Lambda_{1}$-gauge. 
The restrictions imposed on $\omega$ originate then
from the $\sqrt{\ }$ function which enters $\psi_{1}(r)$.  
This completes the presentation of the generalized Kruskal scheme.

\bigskip
In this paper, although restraining ourselves to solely deal
with the geometrical and the topological aspects, we have 
serendipitously challenged the fundamental formula which
governs the quantum theory of black holes.
The Hawking-Gibbons formula, say $\displaystyle{\omega(0)=
\frac{1}{4m}}$ for the prototype Schwarzschild black hole,
is viewed as the tip of an iceberg, the $k\to 0$ limit of a
full class of dispersion relations $\omega(\xi k)$.
On the technical side we have first shown that, for globally
non-trivial (living in $M_{5}\times S_{1}$) $k$-embeddings,
Kasner-Fronsdal extendibility can be achieved for apparently
forbidden frequencies.
Then, motivated by the fact that extendibility actually means
the removal of conic singularities in the (Euclidean) Kruskal
language, we have derived the corresponding Kruskal-like sheets
as slices in some Kaluza-Klein background. 
We are partially aware of the potential impact the present
work may have on black hole physics.
In fact, preliminary results already suggest a discrete
(reflecting the twistor flux quantization) quantum family of
classically degenerate black holes.

\acknowledgments
It is our pleasure to thank Prof. E. Gedalin and T. Shwartz for
stimulating discussions.

\begin{figure}
\caption{The extendibility allowed region $\omega_{1}(k)\le\omega
\le\omega_{2}(k)$. Notice the Hawking-Gibbons limit $\displaystyle
{\omega_{1,2}(0)={1\over{4M}}}$.}
\end{figure}

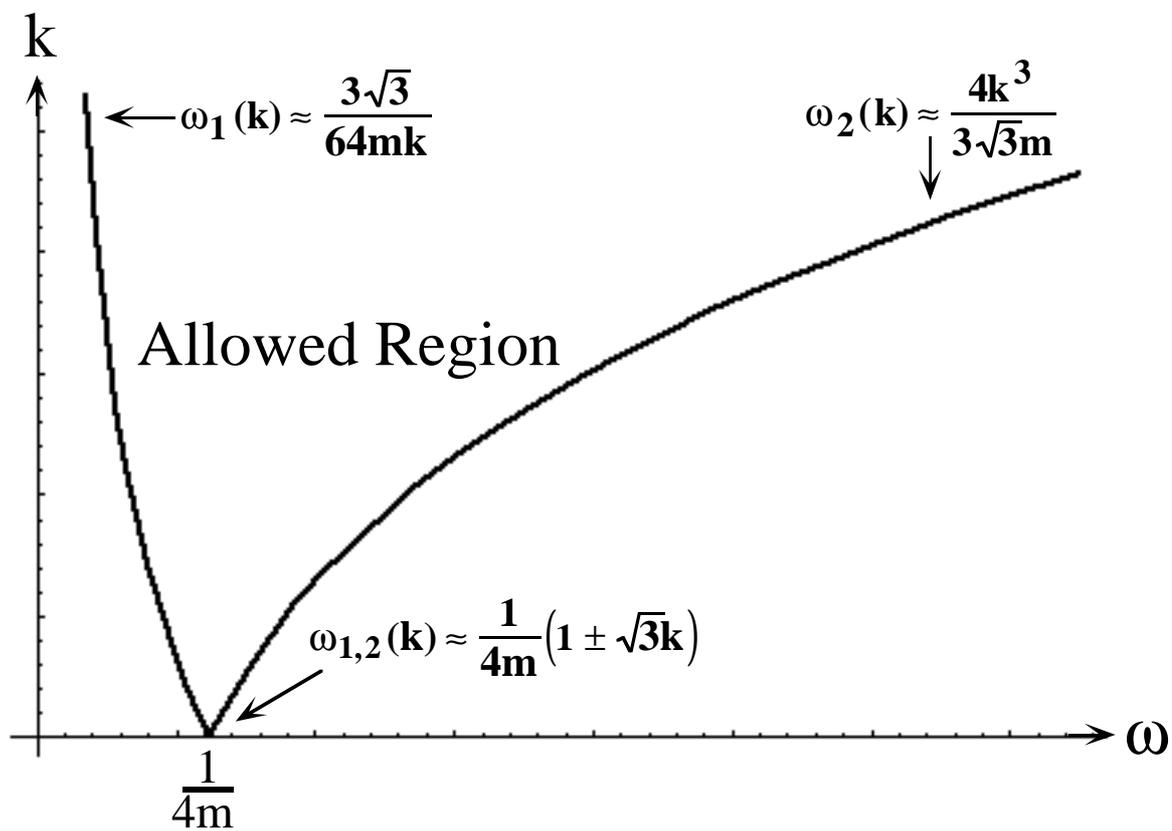
\end{document}